# First results from the PARSE.Insight project: HEP survey on data preservation, re-use and (open) access[1]

Contribution to the First Workshop on Data Preservation and Long-Term Analysis in High-Energy Physics, DESY, Hamburg, Germany, January 26th-28th 2009[2]

*André Holzner[a], Peter Igo-Kemenes[a,b] and Salvatore Mele[a]*

[a] CERN, 1211, Geneva 23, Switzerland.
[b] Gjøvik University College, Po.box 191 Teknologivn. 22, 2802 Gjøvik, Norway.

**Abstract**

There is growing interest in the issues of preservation and re-use of the records of science, in the "digital era". The aim of the PARSE.Insight project, partly financed by the European Commission under the Seventh Framework Program, is twofold: to provide an assessment of the current activities, trends and risks in the field of digital preservation of scientific results, from primary data to published articles; to inform the design of the preservation layer of an emerging e-Infrastructure for e-Science. CERN, as a partner of the PARSE.Insight consortium, is performing an in-depth case study on data preservation, re-use and (open) access within the High-Energy Physics (HEP) community. The first results of this large-scale survey of the attitudes and concerns of HEP scientists are presented. The survey reveals the widespread opinion that data preservation is "very important" to "crucial". At the same time, it also highlights the chronic lack of resources and infrastructure to tackle this issue, as well as deeply-rooted concerns on the access to, and the understanding of, preserved data in future analyses.

## Background

Key stakeholders in the arena of digital preservation of scientific records in Europe and beyond have federated in the *Alliance for Permanent Access* which counts among its members major research organizations, national libraries, publishers and partners involved in digital preservation [1]. CERN is a member of the *Alliance* as a natural extension of its Open Access activities.

[1] This work is partly supported by the European Commission through the PARSE.Insight project (RI-223758).
[2] http://indico.cern.ch/conferenceDisplay.py?confId=42722.

In recent years, the European Commission has played a crucial role in the support of pan-European connectivity, through the GÉANT initiative [2]; the development of Grid computing, through the EGEE suite of projects [3]; the development of digital repositories for scientific artifacts through a variety of projects [4]. The European Commission, as a part of its global vision for e-Science, is now turning its interest to the preservation of scientific artifacts, from primary data to publications, and called for proposals for surveying the status of the field. Members of the *Alliance*, among which CERN, answered the call presenting the PARSE.Insight project which was retained for funding.

The PARSE.Insight project [5] aims to generate insight into current trends regarding the **P**ermanent **A**ccess to the **R**ecords of **S**cience in **E**urope. Its participant represents research institutions, funding agencies, national libraries, commercial publishers and other partners involved in digital preservation. PARSE.Insight, active from March 2008 to February 2010, has performed a large-scale survey of the attitudes of funding agencies, researchers, publishers and libraries towards digital preservation; this is complemented by a number of in-depth case studies in disciplines known for their pioneering approaches to issues in scholarly communication. One of these disciplines is High-Energy Physics.

HEP presents a particularly complex case in the debate on digital preservation, due to the huge amount and outstanding complexity of the data generated by present-day accelerator facilities; it may thus be regarded as a "worst case scenario" for digital preservation.

PARSE.Insight in general, and the HEP case study in particular, is not a study on technical solutions but rather on "soft" issues such as motivations *vs.* concerns, threats *vs.* opportunities, wishes *vs.* obstacles. These are the issues on which the European Commission needs information, to base its evolving strategy for e-Infrastructures in e-Scence. In this sense, the PARSE.Insight project and the workshops on Data Preservation and Long-Term Analysis in HEP are complementary, and have the opportunity to produce together a global picture for the future of data preservation, (open) access and re-use of HEP data.

## The PARSE.Insight HEP case study

The central component of the case study of the PARSE.Insight project about data preservation, (open) access and re-use in HEP is an online survey. It was launched in October 2008 and was running for three months. It was advertised through the mailing lists of large experimental collaborations and through a link on the SPIRES web page. A large fraction of the HEP community (which is estimated to include about 20,000 active physicists) was thus reached, yielding 1197 responses: 883 by experimental physicists and 314 by theoretical physicists.

The questions were grouped in five categories:

- Demographics;
- Perception of the importance of data preservation and its motivation;
- Which kind of information should be preserved: granularity and level of abstraction;
- When, how and where should the data be preserved;
- Concern raised by (open) access to preserved HEP data.

Most questions offered multiple-choice answers. In addition several allowed free-text answers. Many respondents made use of this opportunity and 2550 free-text answers are now being evaluated. Many respondents expressed the wish to be interviewed on specific, related, issues; these interviews will be taken up soon.

This contribution presents the preliminary results of a first analysis of the multiple-choice responses, and is restricted to a few key messages relevant for this series of workshops. A graphical summary of most responses is available in the slides presented at the workshop[3].

## Demographics

The distribution by country of the respondents reflects approximately that of the active physicists in the field [6]: 41.1% come from countries of the European Union, 23.1% from the United States, and 23.1% from the rest of the world, while 12.7% spend most of their working time at CERN which was presented as an additional "country" in this study. Here and in the following the percentages are calculated with respect to the total number of answers received for a particular question. Cases where a question was left unanswered are discarded from the calculation.

The distribution of respondents in their career path is rather flat, with 22.9% Ph.D. students, 23.3% post-doctoral fellows, 28.0% researchers with permanent positions, and 25.7% professors.

Experimental physicists were also asked to specify the large projects in which they are or were involved. This question allowed multiple answers. The LHC experiments were indicated by 70.3% of the respondents; LEP by 16.3%; CDF or D0 by 19.8%; H1 or ZEUS by 9.9%, BaBar, BELLE or CLEO by 19.1%. In addition, 12.4% of the respondents are or were involved in neutrino programs; 6.2% in kaon programs; 4.0% in heavy-ion physics. In general 17.2% are or were active at fixed-target experiments represent 17.2% and 9.9% in future projects such as ILC,

[3] http://indico.cern.ch/materialDisplay.py?contribId=15&sessionId=6&materialId=slides&confId=42722.

SLHC or CLIC. 15.7% of respondents are or were active in other programs not covered in the list.

The respondents are therefore representative of the entire HEP community: geographically, by career status and age group, and by covering a large spectrum of HEP experiments.

## The importance of data preservation

Figure 1 shows the distribution of the answers to the crucial question "How important is the issue of data preservation?". The answers are given separately for theorists (top/blue) and experimentalists (bottom/green). It is remarkable that about 68.9% of the respondents perceive preservation as "very important" or even "crucial". The distributions are the same if the respondents are divided in "age groups" (less or more than 5 years of experience in HEP).

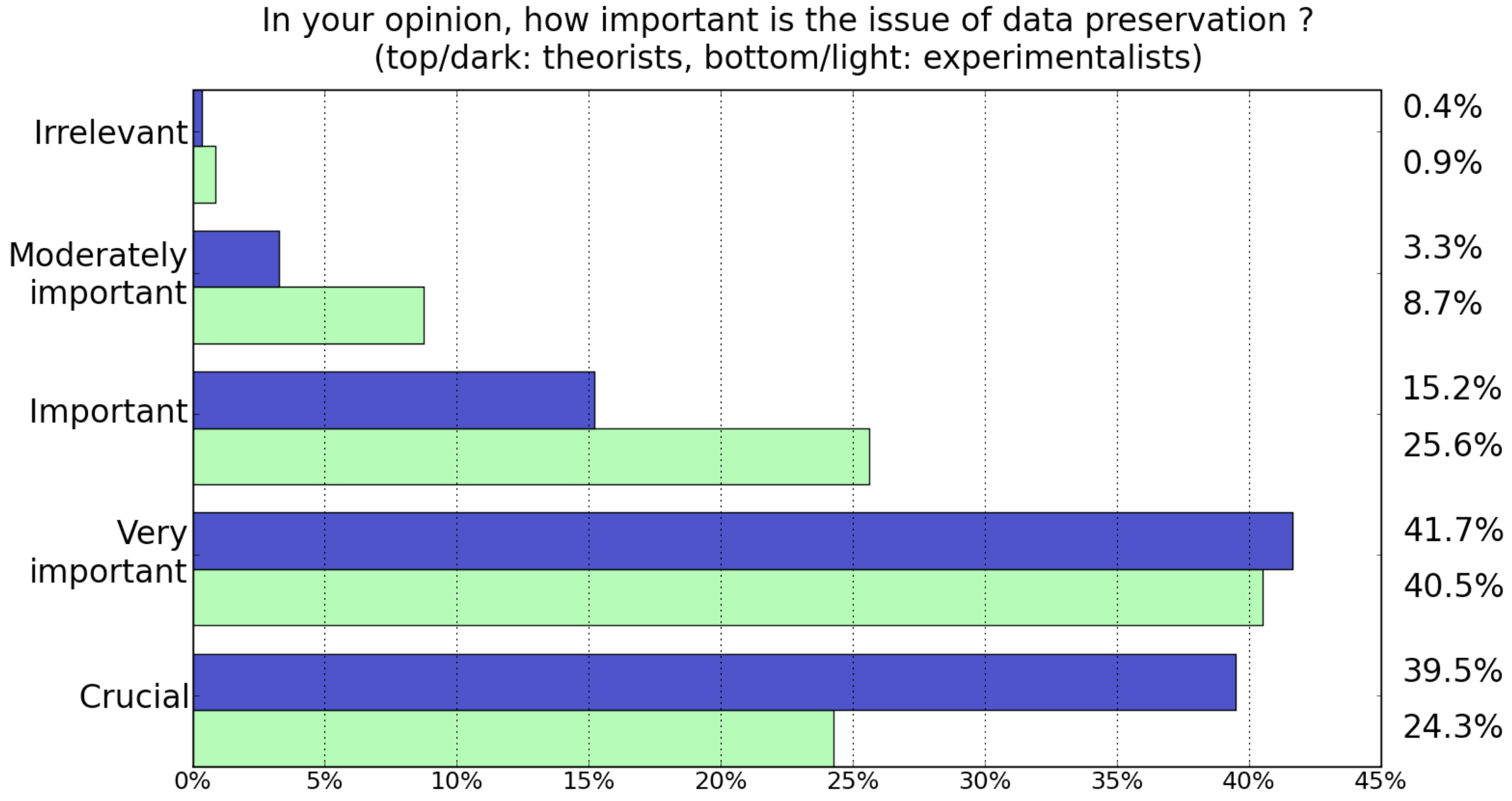

*Figure 1: The perceived importance of data preservation*

It is interesting to divide the sample in affiliations to major experimental projects. Preservation is "very important or crucial" for 56.3% of participants to LEP experiments, 65.9% for HERA and 64.3% for LHC experiments. Remarkably, this percentage is only 50.0% for Tevatron experiments. This relatively low perception of importance might reflect the thinking that the future LHC collider program is the straight continuation of the Tevatron program covering both similar fields of physics and possibly reducing interest in preserving Tevatron data at the onset of the LHC era.

Four reasons for preservation were offered to the respondents, who were asked to indicate for each of them a level of importance. A level of "very important" or "crucial" was indicated for

- Use of preserved data for future independent checks of results: 60.8%;
- re-analysis of preserved data to test future theories: 73.6%;
- combination of preserved data with future data: 62.8%;
- use of preserved data for teaching or outreach: 27.2%.

Each of the first three use cases is perceived to be more important by theorists than by experimentalists. This difference is striking for the use of preserved data to allow independent checks of results this is particularly true for the first use case, deemed "very important" or "crucial" by 80.8% for theorists but only 54.1% of experimentalists. The relatively low interest in re-using preserved data for teaching or outreach is somewhat surprising.

The scientific case for data preservation and re-use is epitomized by the fact that: 53.7% of the theorists and 43.8% of the experimentalists think that access to data from past experiments could have improved their scientific results. At the same time, 46.2% of respondents think that important HEP data have been lost in the past.

## What should be preserved, when, and where?

Figure 2 summarizes the answers to the question regarding the level of abstraction at which data should be preserved. Moving downwards from the top, the complexity of the data is increasing while the level of abstraction decreases. The distribution is remarkably flat. In the light of the high complexity of the problem of re-using older data, it is surprising that as many as 66.6% or respondents would preserve event-by-event information and 45.1% even raw data.

Regarding the time at which the data should be made available for preservation and consequent re-use by others, 6.8% think that this should happen immediately after the data have been recorded, 31.9% once the analysis is completed and the results published, 21.4% would impose some "embargo time" following the publication and 39.3% would do it only at the very end of the project. Young physicists with less than 5 years experience in the field are more inclined to make their data available immediately after recording while more experienced researchers show a clear preference for providing access only once the project is finished.

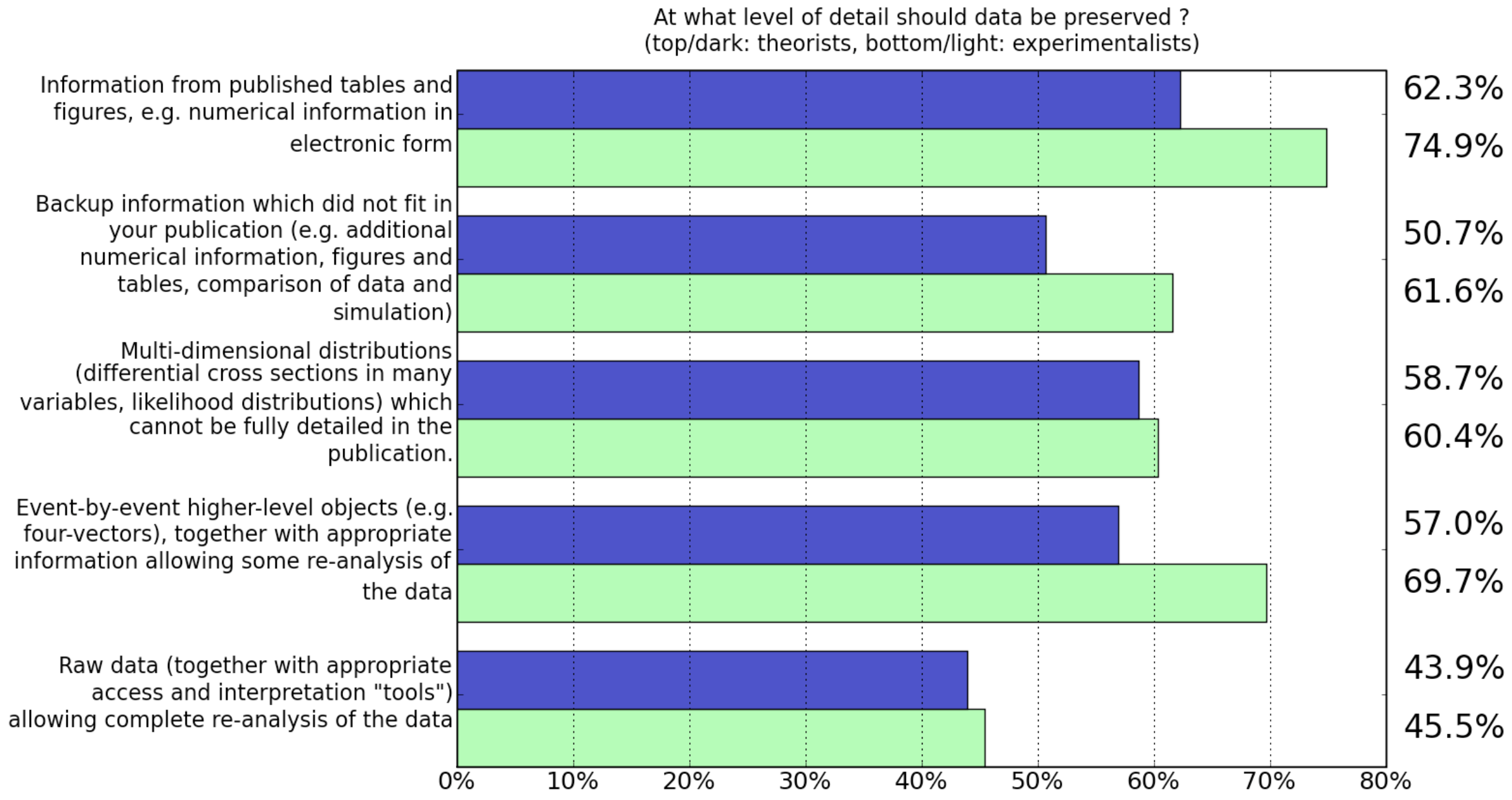

*Figure 2: At which level should data be preserved?*

The overwhelming majority of respondents (68.4%) would like data to be preserved at a "neutral" platform such the equivalent of ADS, arXiv, CDS or SPIRES, adapted to the preservation of and access to data. This preference is even stronger in the case of theorists (85.6%). Storage at a site connected to the experiment or laboratory where the data have been produced is the second choice, with a stronger preference from the side of the experimentalists. This might reflect a certain level of concern of theorists regarding the stability of data when hosted by the producers themselves. Platforms managed by journal publishers receive a very low preference.

Respondents were asked to qualify the attributes of such a preservation platform. While the analysis of these free-text answers is ongoing, Figure 3 presents the corresponding tag cloud. Access, documentation, openness are desires clearly expressed by the community.

Almost all respondents (94.3%) think that the additional effort needed for preparing data for preservation in a re-usable form is substantial (more than 1% of the overall effort invested in producing and analyzing the data) whereas 43.0% think that the supplementary effort is more than 10%. These percentages are independent of the experimental facilities where respondents work. This finding is not surprising and confirms the important financial and person-power implications of a large-scale preservation program in HEP. This situation is made more complex by the timing that researchers indicate as crucial for the success of a data preservation program: 41.4% of the respondents think that the effort towards data preservation has to be deployed concurrently to data taking, while 28.1% think that preservation should be prepared even before the actual data taking. It is remarkable that participants to the LHC experiments strongly indicate that data preservation should be addressed *before* data taking, which means *now*!

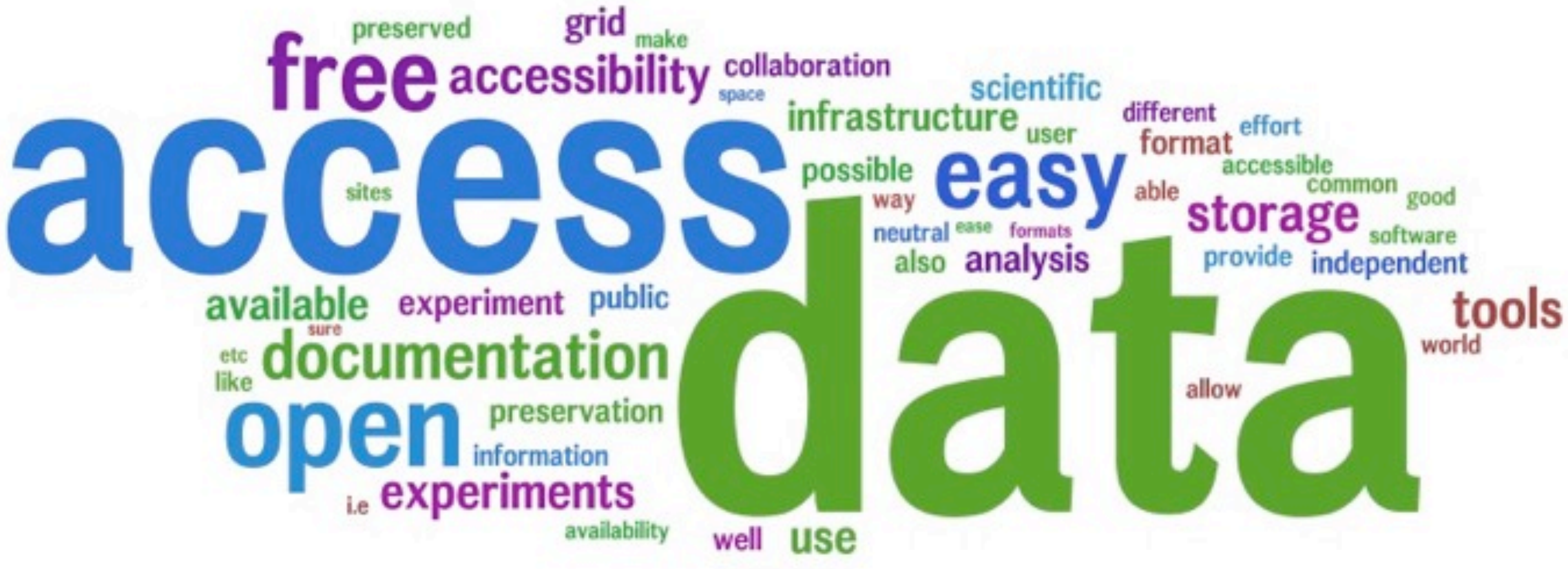

*Figure 3: Main attributes of an e-Infrastructure for HEP data preservation and access*

Against this background of positive attitude towards data preservation and acknowledgement of the large challenge it present, it is a sobering to find that only 16.1% of respondents thinks that their experiment/collaboration/organization is able to produce the required financial effort and person-power to tackle this issue, 6.5% think that this is not the case; the large majority of respondents, 77.4%, just does not know, showing a clear gap between awareness and action.

## Concerns

The survey also aimed to quantify issues clouding the potential of third-party re-use of preserved data. Two potential areas of concern emerge: the sharing of credit and responsibilities between the producers and re-users of the data, and the validity of the results derived from the analysis of preserved data.

A relatively low number of respondents (24.0%) are “very concerned” or “gravely concerned” that preserved data could be used without giving proper credit to the authors; of these, young scientists (<5 years in the field) seem to be only slightly more concerned than more experienced researchers.

On the other hand, 45.0% of the respondents are "very concerned" or "gravely concerned" that data re-use may in general lead to an inflation of incorrect results. Experimentalists are by far more concerned (51.3%) than theorists (29.0%). At the same time, respondents are also concerned by producing themselves uncorrected results by misinterpreting preserved data. As many as 53.2% are "very concerned" or "gravely concerned" about this possibility and, of these, experimentalists are again more concerned than theorists.

## Final remarks

The analysis of the answers received in response to the online HEP survey on data preservation, re-use and (open) access has only started. However, a clear picture emerges, of a community which is both engaged and concerned by the issue of data preservation, clearly acknowledging the need for urgent action and the vast investment this action requires. The forthcoming analysis of free-text answers and the forthcoming interviews, in correlation with the demographics of the respondents, will help producing a community-wide picture and contribute to formulating strategy for data preservation in HEP, in synergy with the outcomes of the present series of workshops.